\documentclass[12pt]{article}
\usepackage{epsfig}
\def\be{\begin{equation}}
\def\ee{\end{equation}}
\def\bea{\begin{eqnarray}}
\def\eea{\end{eqnarray}}
\usepackage{graphicx}

\catcode`\@=11
\def\lsim{\mathrel{\mathpalette\@versim<}}
\def\gsim{\mathrel{\mathpalette\@versim>}}
\def\@versim#1#2{\vcenter{\offinterlineskip
\ialign{$\m@th#1\hfil##\hfil$\crcr#2\crcr\sim\crcr } }}
\catcode`\@=12

\catcode`@=12
\begin{document}
\thispagestyle{empty}
\begin{flushright}
UCRHEP-T553\\
June 2015\
\end{flushright}
\vspace{0.4in}
\begin{center}
{\LARGE \bf Radiative Mixing of the One Higgs Boson\\ 
and Emergent Self-Interacting Dark Matter\\}
\vspace{0.8in}
{\bf Ernest Ma\\}
\vspace{0.2in}
{\sl Physics \& Astronomy Department and Graduate Division,\\ 
University of California, Riverside, California 92521, USA\\}
\vspace{0.2in}
{\sl HKUST Jockey Club Institute for Advanced Study,\\ 
Hong Kong University of Science and Technology, Hong Kong, China\\}
\end{center}
\vspace{0.8in}

\begin{abstract}\
In all scalar extensions of the standard model of particle interactions, 
the one Higgs boson responsible for electroweak symmetry breaking always 
mixes with other neutral scalars at tree level unless a symmetry prevents 
it.  An unexplored important option is that the mixing may be radiative, 
and thus guaranteed to be small.  Two first such examples are discussed. 
One is based on the soft breaking of the discrete symmetry $Z_3$. The 
other starts with the non-Abelian discrete symmetry $A_4$ which is then softly 
broken to $Z_3$, and results in the emergence of an interesting dark-matter 
candidate together with a light mediator for the dark matter to have its own 
long-range interaction.
\end{abstract}

\newpage
\baselineskip 24pt

The standard model (SM) of particle interactions requires only one scalar 
doublet $\Phi = (\phi^+,\phi^0)$ which breaks the electroweak gauge symmetry 
$SU(2)_L \times U(1)_Y$ spontaneously to electromagnetic $U(1)_Q$ with 
$\langle \phi^0 \rangle = v = 174$ GeV.  It predicts just one physical 
Higgs boson $h$, whose properties match well with the 125 GeV particle 
discovered~\cite{atlas12,cms12} at the Large Hadron Collider (LHC) in 2012. 
If there are other neutral scalars at the electroweak scale, they are 
expected to mix with $h$ at tree level, so that the observed 125 GeV 
particle should not be purely $h$.  If future more precise measurments 
fail to see deviations, then a theoretical understanding could be that 
a symmetry exists which distinguishes $h$ from the other scalars. 
For example, a $Z_2$ symmetry may exist under which a second scalar 
doublet $(\eta^+,\eta^0)$ is odd~\cite{dm78} and all SM particles are even.
This is useful for having a viable dark-matter candidate~\cite{atyy14}. 
It also enables the simplest one-loop (scotogenic) model~\cite{m06} of 
radiative neutrino mass through dark matter, if three neutral singlet 
fermions $N_{1,2,3}$ are also added which are odd under $Z_2$, as shown 
in Fig.~1. 
\begin{figure}[htb]
\vspace*{-3cm}
\hspace*{-3cm}
\includegraphics[scale=1.0]{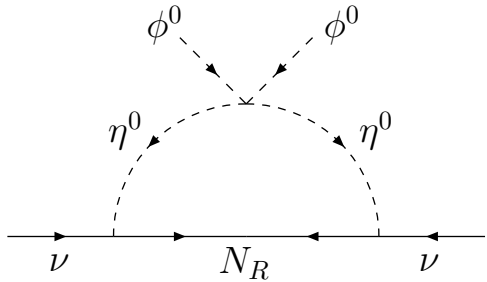}
\vspace*{-21.5cm}
\caption{One-loop $Z_2$ scotogenic neutrino mass.}
\end{figure}
Note that this dark $Z_2$ symmetry may be derived from lepton 
parity~\cite{m15}.

Another important new development in the understanding of dark matter 
is the possibility that it has long-range self-interactions~\cite{t14}.  
This may be an elegant solution to two existing discrepancies in 
astrophysical observations: (1) central density profiles of dwarf 
galaxies are flatter (core) than predicted (cusp); (2) observed number 
of satellites in the Milky Way is much smaller than predicted.  Although 
(2) is less of a problem with the recent discovery of faint satellites, 
there is also (3) the new related problem of predicted massive subhalos 
which are not observed.  From a particle theory perspective, if the light 
mediator of dark matter is a scalar boson, then it is difficult to construct a 
theory such that it does not mix substantially with $h$.  In this paper it 
will be shown how a dark-matter candidate could emerge together with its 
mediator, such that its mixing with $h$ is one-loop suppressed.  This will 
be discussed later as part of my second example.

If an exact dark symmetry is imposed, then any mixing of $h$ with other 
neutral scalars in the dark sector would be forbidden.  However, if this 
symmetry is softly broken, it will result in nonzero mixing, many examples 
of which exist.  In the above scotogenic model, if the quadratic term 
$m_{12}^2 \Phi^\dagger \eta + H.c.$ is added thus breaking $Z_2$ softly, 
a nonzero vacuum expectation value $\langle \eta^0 \rangle$ would be 
induced~\cite{m01}. This invalidates $\eta^0$ as a dark-matter candidate.  
However $\langle \eta^0 \rangle$ could be naturally small as an explanation 
of the smallness of neutrino masses~\cite{m01}.  At the same time, $h$ 
mixes with $\eta^0$ at tree level as expected.  To obtain radiative mixing 
which has never been discussed before, my first example is the soft breaking 
of a $Z_3$ symmetry which would lead to the one-loop mixing of $h$ with two 
other scalars with zero vacuum expectation values. 

Consider the addition of a complex neutral scalar singlet $\chi$ to the SM, 
transforming as $\omega = \exp(2\pi i/3)$ under $Z_3$.  Let the scalar 
potential of $\Phi$ and $\chi$ be given by
\begin{eqnarray}
V &=& \mu^2 \Phi^\dagger \Phi + m_1^2 \chi^\dagger \chi + 
{1 \over 2} m_2^2 \chi^2 + {1 \over 2} (m^*_2)^2 (\chi^\dagger)^2 + 
{1 \over 3} f \chi^3 + {1 \over 3} f^* (\chi^\dagger)^3 \nonumber \\  
&+& {1 \over 2} \lambda_1 (\Phi^\dagger \Phi)^2 + {1 \over 2} \lambda_2 
(\chi^\dagger \chi)^2 + \lambda_3 (\chi^\dagger \chi) (\Phi^\dagger \Phi),
\end{eqnarray}
where $Z_3$ is softly broken only by the quadratic $m_2^2$ and $(m_2^*)^2$ 
terms.  The phase of $\chi$ may be redefined to render $m_2^2$ real, but 
then $f$ must remain complex in general.  As $\phi^0$ acquires a nonzero 
vacuum expectation value $\langle \phi^0 \rangle = v$ so that $m_h^2 = 
2 \lambda_1 v^2$, the $2 \times 2$ mass-squared matrix for $\chi = (\chi_R + 
i \chi_I)/\sqrt{2}$ becomes
\begin{equation}
{\cal M}_\chi^2 = \pmatrix{m_1^2 + \lambda_3 v^2 + m_2^2 & 0 \cr 0 & 
m_1^2 + \lambda_3 v^2 - m_2^2} = \pmatrix{m_R^2 & 0 \cr 0 & m_I^2}.
\end{equation}
The interaction Lagrangian for $h, \chi_R, \chi_I$ is then
\begin{eqnarray}
-{\cal L}_{int} &=& {1 \over 2} m_h^2 h^2 + {1 \over \sqrt{2}} \lambda_1 v h^3 
+ {1 \over 8} \lambda_1 h^4 + {\lambda_3 v \over \sqrt{2}} h (\chi_R^2 + 
\chi_I^2) + {\lambda_3 \over 4} h^2 (\chi_R^2 + \chi_I^2) \nonumber \\ 
&+& {1 \over 2} m_R^2 \chi_R^2 + {1 \over 2} m_I^2 \chi_I^2 + {1 \over 8} 
\lambda_2 (\chi_R^2 + \chi_I^2)^2 \nonumber \\ &+& {f_R \over 3 \sqrt{2}} 
\chi_R^3 - {f_I \over \sqrt{2}} \chi_R^2 \chi_I - {f_R \over \sqrt{2}} 
\chi_R \chi_I^2 + {f_I \over 3 \sqrt{2}} \chi_I^3.
\end{eqnarray}
It is clear that $h$ does not mix with $\chi_R$ or $\chi_I$ at tree level. 
However there will be radiative mixing in one loop as shown in Fig.~2.
\begin{figure}[htb]
\vspace*{-4cm}
\hspace*{-3cm}
\includegraphics[scale=1.0]{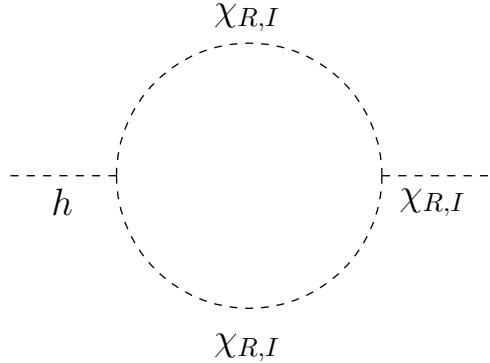}
\vspace*{-19.5cm}
\caption{Radiative mixing of $h$ with $\chi_{R,I}$.}
\end{figure}
The $h \chi_R$ mixing is given by
\begin{equation}
2i \lambda_3 f_R v \int {d^4 k \over (2 \pi)^4} \left[ 
{1 \over (k^2 - m_R^2)^2} - {1 \over (k^2 - m_I^2)^2}
\right] = {\lambda_3 f_R v \over 8 \pi^2} \ln {m_R^2 \over m_I^2}.
\end{equation}
Similarly, the $h \chi_I$ mixing is obtained with $f_R$ replaced by $f_I$ 
and $m_{R,I}$ by $m_{I,R}$.  This new phenomenon allows the 125 GeV 
particle to be essentially $h$ and yet the scalar sector is enriched 
with other possible consequences different from that of an exact symmetry 
such as dark parity.

A useful application of the above scenario is in connection with the simplest 
model~\cite{sz85} of dark matter (DM), i.e. that of a real neutral singlet 
$s$, which is odd under dark $Z_2$.  It has been shown recently~\cite{fpu15} 
that most of its parameter space has been ruled out by the LUX direct-search 
experiment for dark matter~\cite{lux14}.  The tension comes from the 
requirement that the $ss$ annihilation cross section to be of the correct 
magnitude to account for the observed DM relic density of the 
Universe, but its interaction with nuclei through $h$ exchange to be below 
the LUX bound.  To satisfy the latter, the former becomes too small, hence 
the $s$ relic abundance would exceed what is observed.  With the addition 
of $\chi$, the allowed $ss \chi^\dagger \chi$ interaction would add to the 
$ss$ annihilation cross section, but would not contribute to the 
direct-search constraint.  In this way the parameter space for $s$ dark 
matter opens up to $m_s > m_\chi$.  This solution also applies to 
the recently proposed $A_4$ model~\cite{m15-1} of neutrino mass, where 
$s_{1,2,3} \sim \underline{3}$ and the lightest is dark matter.  Here 
$\chi \sim \underline{1}'$ of $A_4$.

My second example of radiative mixing of $h$ and another scalar is based 
on the breaking of $A_4$ to $Z_3$.  Consider now three real neutral scalar 
singlets $\chi_{1,2,3} \sim \underline{3}$ of $A_4$.  Let their scalar 
potential with $\Phi$ be given by
\begin{eqnarray}
V &=& \mu^2 \Phi^\dagger \Phi + {1 \over 2} m_1^2 (\chi_1^2 + \chi_2^2 + 
\chi_3^2) + m_2^2 (\chi_1 \chi_2 + \chi_2 \chi_3 + \chi_3 \chi_1) 
+ f \chi_1 \chi_2 \chi_3 + {1 \over 2} \lambda_1 (\Phi^\dagger \Phi)^2 
\nonumber \\  &+&  {1 \over 4} \lambda_2 
(\chi_1^2 + \chi_2^2 + \chi_3^2)^2 + {1 \over 4} \lambda_3 (\chi_1^2 \chi_2^2 
+ \chi_2^2 \chi_3^2 + \chi_3^2 \chi_1^2) + {1 \over 2} \lambda_4 
(\chi_1^2 + \chi_2^2 + \chi_3^2)(\Phi^\dagger \Phi),
\end{eqnarray}
where $A_4$ is broken softly to $Z_3$ by the $m_2^2$ term.  Note that the 
trilinear $f$ term means that $\chi_{1,2,3}$ do not have the dark parity 
of the similar $s_{1,2,3}$ scalars discussed previously.  Let~\cite{mr01}
\begin{equation}
\pmatrix{\chi_0 \cr \chi \cr \chi^\dagger} = 
{1 \over \sqrt{3}} \pmatrix{1 & 1 & 1 \cr 1 & \omega & \omega^2 \cr 
1 & \omega^2 & \omega} 
\pmatrix{\chi_1 \cr \chi_2 \cr \chi_3},
\end{equation}
then the real $\chi_0$ and complex $\chi$ are mass eigenstates with 
$m_0^2 = m_1^2 + \lambda_4 v^2 + 2 m_2^2$ and 
$m_\chi^2 = m_1^2 + \lambda_4 v^2 - m_2^2$.  Under the residual $Z_3$ symmetry, 
$\chi_0 \sim 1$, $\chi \sim \omega$, and $\chi^\dagger \sim \omega^2$. 
The interaction Lagrangian for $h, \chi_0, \chi$ is then
\begin{eqnarray}
-{\cal L}_{int} &=& {1 \over 2} m_h^2 h^2 + {1 \over \sqrt{2}} \lambda_1 v h^3 
+ {1 \over 8} \lambda_1 h^4 + {\lambda_4 v \over \sqrt{2}} h (\chi_0^2 + 
2\chi \chi^\dagger) + {\lambda_4 \over 4} h^2 (\chi_0^2 + 2\chi \chi^\dagger) 
\nonumber \\ 
&+& {1 \over 2} m_0^2 \chi_0^2 + m_\chi^2 \chi \chi^\dagger + {1 \over 4} 
\lambda_2 (\chi_0^2 + 2 \chi \chi^\dagger)^2 + {f \over 3 \sqrt{3}} 
[\chi_0^3 + \chi^3 + (\chi^\dagger)^3 - 3 \chi_0 \chi \chi^\dagger] 
\nonumber \\ 
&+&  {1 \over 12} \lambda_3 [\chi_0^4 - 2 \chi_0 \chi^3 
- 2 \chi_0 (\chi^\dagger)^3 + 3 (\chi \chi^\dagger)^2].
\end{eqnarray}
Again there is no tree-level mixing for $h$, but radiative mixing occurs 
between $h$ and $\chi_0$ as shown in Fig.~3.
\begin{figure}[htb]
\vspace*{-4cm}
\hspace*{-3cm}
\includegraphics[scale=1.0]{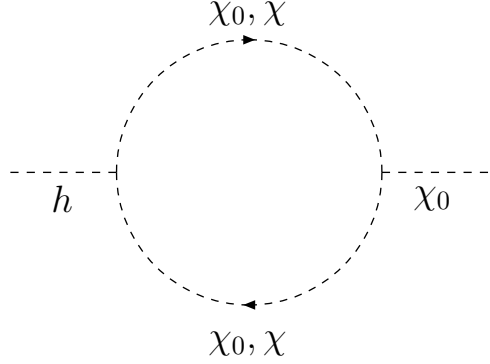}
\vspace*{-19.5cm}
\caption{Radiative mixing of $h$ with $\chi_0$.}
\end{figure}
This mixing is given by
\begin{equation}
m^2_{h \chi} = \sqrt{(2/3)} i \lambda_4 f v \int {d^4 k \over (2 \pi)^4} \left[ 
{1 \over (k^2 - m_0^2)^2} - {1 \over (k^2 - m_\chi^2)^2}
\right] = {\lambda_4 f v \over 8 \sqrt{6} \pi^2} \ln {m_0^2 \over m_\chi^2}.
\end{equation}
Hence $\chi_0$ would decay to SM particles through its mixing with $h$. 
On the other hand, $\chi$ itself is absolutely stable and becomes a 
suitable DM candidate.  It is an example of $Z_3$ dark 
matter~\cite{bkpr13,kt14}, which may also be connected with the two-loop 
generation of neutrino mass~\cite{m08}.

The important thing to notice in this model is the natural emergence of 
$\chi$ and $\chi_0$ as the result of $A_4$ breaking to $Z_3$ with their 
accompanying interactions as shown in Eq.~(7).  Whereas $\chi \sim \omega$ 
under $Z_3$ emerges as a dark-matter candidate, $\chi_0 \sim 1$ emerges 
automatically as its mediator.  In other words, all the ingredients of 
self-interacting dark matter appear at once.  Note that this desirable 
outcome is the result of $A_4$ breaking to $Z_3$.  If only $Z_3$ is assumed 
in the beginning, then $\chi_0 \sim 1$ would have a substantial tree-level 
mixing with $h$. Since the mediator in self-interacting matter is required 
to be light, say 10 MeV, the fact that its mixing with $h$ is one-loop 
suppressed by Eq.~(8) is an important desirable feature of this model.

Since $\chi_0$ is presumably in thermal equilibrium with all the SM particles, 
it would be overproduced in the early Universe if it is stable.  To avoid 
any conflict with the very successful predictions of Big Bang Nucleosynthesis 
(BBN), its decay lifetime should be less than about one second.  For 
$m_{\chi_0} = 10$ MeV, $\chi_0$ would decay to $e^- e^+$ or to $\gamma \gamma$ 
through its mixing with $h$.  In contrast to the case of 
$h$ decay, the former is dominant because the latter lacks the extra 
enhancement of $m_h^2/m_{\chi_0}^2$.   The $\chi_0 \to e^- e^+$ decay rate 
is easily calculated to be
\begin{equation}
\Gamma (\chi_0 \to e^- e^+) = {m_{\chi_0} m_e^2 \over 16 \pi v^2} \left( 
{m_{h \chi}^2 \over m_h^2} \right)^2.
\end{equation}
For $m_{\chi_0} = 10$ MeV, $m_e = 0.511$ MeV, $v = 174$ GeV, the mixing of 
$\chi_0$ with $h$ is constrained by $(m_{h \chi}^2/m_h^2) > 2 \times 10^{-5}$ if 
$\Gamma^{-1} < 1$ sec.  Using Eq.~(8), this translates to 
\begin{equation}
\lambda_4 f > 0.02~{\rm GeV}.
\end{equation}  
To be successful as self-interacting dark matter, the trilinear scalar 
interaction $(f/\sqrt{3})\chi_0 \chi \chi^\dagger$ in Eq.~(7) should have 
a magnitude of order the weak scale.  Hence the above condition is easily 
satisfied.  On the other hand, the $2 \times 2$ mass-squared matrix 
spanning $h$ and $\chi_0$ is given by
\begin{equation}
{\cal M}^2 = \pmatrix{m_h^2 & m^2_{h \chi} \cr m^2_{h \chi} & m_0^2}.
\end{equation}
For $m_0^2 << m_h^2$, the $h - \chi_0$ mixing is simply $m^2_{h \chi}/m_h^2$ 
but the physical mass of $\chi_0$ is also modified, i.e.
\begin{equation}
m^2_{\chi_0} = m_0^2 - m^4_{h \chi}/m_h^2.
\end{equation}
For $m^2_{h \chi}/m_h^2 = 2 \times 10^{-5}$, the correction is 
$(2 \times 10^{-5})^2 (125~{\rm GeV})^2 \simeq 6~{\rm MeV}^2$.  If $m_0 = 10$ 
MeV, $m_{\chi_0}$ is reduced only to 9.7 MeV .  This is of course 
acceptable, but if $m^2_{h \chi}/m_h^2$ is too large, $m_0$ must be 
delicately adjusted to obtain $m_{\chi_0} \sim 10$ MeV.  To avoid too 
much cancellation, $m^2_{h \chi}/m_h^2 < 10^{-4}$ is a reasonable 
condition, for which
\begin{equation}
\lambda_4 f < 0.1~{\rm GeV}.
\end{equation}
Given that $f \sim 100$ GeV is required, $\lambda_4 < 10^{-3}$ is implied.

The relic abundance of $\chi$ is determined by its annihilation cross 
section as it decouples from other matter in the early Universe.  The main 
interactions are shown in Fig.~4.
\begin{figure}[htb]
\vspace*{-3cm}
\hspace*{-3cm}
\includegraphics[scale=1.0]{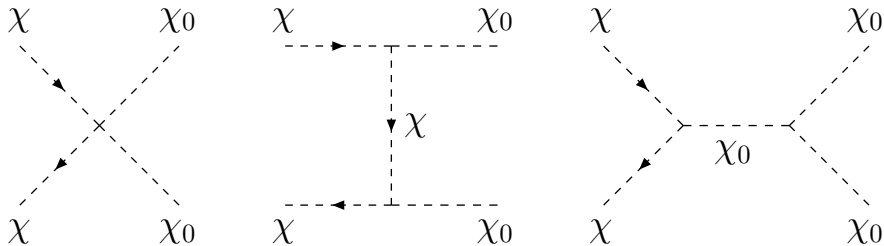}
\vspace*{-21.5cm}
\caption{$\chi \chi^\dagger$ annihilation to $\chi_0$ final states.}
\end{figure}
As a result,
\begin{equation}
\sigma \times v_{rel} = {1 \over 16 \pi m_\chi^2} \left( \lambda_2 - {f^2 
\over 4 m_\chi^2} \right)^2.
\end{equation}
Setting this equal to the optimal value~\cite{sdb12} of $4.4 \times 10^{-26} 
~{\rm cm}^3~{\rm s}^{-1}$ implied by the observed relic abundance of dark 
matter in the Universe,
\begin{equation}
\lambda_2 - {f^2 \over 4 m_\chi^2} = 0.0435 \left( {m_\chi \over 
100~{\rm GeV}} \right)
\end{equation}
is obtained.  The final-state $\chi_0$ particles decay into $e^- e^+$ pairs 
through which thermal equilibrium with all other SM particles may be 
established.

The elastic scattering of $\chi$ off nuclei proceeds through $h$ exchange. 
For $m_\chi = 100$ GeV, the LUX bound of $10^{-45}~{\rm cm}^2$ 
implies~\cite{fpu15}
\begin{equation}
\lambda_4 < 0.01.
\end{equation}
This means that direct-search experiments will need to improve by at least 
one or two orders of magnitude to see this effect.  The $h \to \chi_0 \chi_0$ 
decay rate is given by
\begin{equation}
\Gamma (h \to \chi_0 \chi_0) = {\lambda_4^2 v^2 \over 16 \pi m_h} = 
\left( {\lambda_4 \over 0.01} \right)^2 0.5~{\rm MeV}.
\end{equation}
This would be a contribution to the Higgs invisible width because the 
lifetime of $\chi_0$ will be certainly long enough for it to escape 
detection at the LHC.

In conclusion, I have discussed the new idea that scalars beyond the SM 
may only mix with the one Higgs boson $h$ (identified as the 125 GeV 
particle discovered at the LHC) in one loop.  Two examples have been 
presented.  One is based on the soft breaking of $Z_3$.  Another is 
based on the soft breaking of $A_4$ to $Z_3$.  In the latter, three real 
neutral scalars $\chi_{1,2,3}$ are added which transform as $\underline{3}$ 
under $A_4$.  With the soft breaking, they are reorganized by Eq.~(6) 
into a real $\chi_0 \sim 1$ and a complex $\chi \sim \omega$ under the 
residual $Z_3$, which is exactly conserved.  Whereas $\chi_0$ mixes with 
$h$ radiatively, $\chi$ is a stable dark-matter candidate, with self 
interactions mediated by $\chi_0$.  In this way, all the ingredients 
necessary for self-interacting dark matter emerge together from the 
softly broken symmetry.  The relic abundance of $\chi$ is determined 
by the $\chi \chi^\dagger \to \chi_0 \chi_0$ annihilation cross section. 
The direct detection of $\chi$ from its elastic scattering off nuclei 
is possible through $h$ exchange.  The light mediator $\chi_0$ decays 
mainly into $e^- e^+$, which may be the source of the observed positron 
excess in some astrophysical experiments~\cite{pamela09,ams13}.

\medskip
This work is supported in part 
by the U.~S.~Department of Energy under Grant No.~DE-SC0008541.

\bibliographystyle{unsrt}

\end{document}